# GAMMA-RAYS FROM $^{nat}$Sn AND $^{nat}$C INDUCED BY FAST NEUTRONS

I. M. Kadenko[1], V. A. Plujko[1], B.M. Bondar[1,2],

O. M. Gorbachenko[1], B. Yu. Leshchenko[1,3], K. M. Solodovnyk[1]

[1]*Nuclear Physics Department, Taras Shevchenko National University, Kyiv, Ukraine*
[2]*All-Ukrainian Center for Radiosurgery of the Clinical Hospital "Feofania", Kyiv, Ukraine*
[3]*National Technical University of Ukraine "Kyiv Polytechnic Institute", Kyiv, Ukraine*

The cross-sections of prompt gamma-ray production from $^{nat}$Sn and $^{nat}$C elements induced by 14.1-MeV neutrons were measured. The time-of-flight technique was used for n-γ discrimination. The experimental results were compared with theoretical calculations performed by Empire 3.2 and Talys 1.6 codes using different models for photon strength function and nuclear level density.

## Introduction

Modern developments of nuclear fusion technologies require new nuclear data and refinements of existing data on DT-neutrons interactions [1]. For radiation damage and gamma-ray shielding calculations, it is necessary to know the cross-sections of (n,xγ) reactions. Apart from practical applications, the gamma-ray spectra can also give information about the nuclear structure, excited states, their population and decay. Furthermore, the comparison of experimental results with corresponding theoretical calculations enables to analyze different theoretical approaches describing nuclear reaction mechanisms and to study a reliability different calculations codes for estimations of the gamma-ray productions.

## Experimental results

The amplitude spectra of gamma-rays were measured in a circular geometry using time-of-flight technique based on pulse neutron generator (NPG-200), designed and manufactured at Nuclear Physics Department of Taras Shevchenko National University of Kyiv. The neutron flow was ~$10^7$ n/s with pulse frequency 7.25 MHz. Investigated samples of $^{nat}$Sn and $^{nat}$C with masses $m_{Sn}$=*3208* g, $m_C$ = *1600* g were irradiated by neutrons with the energy of 14.1±0.2 MeV, the gamma-ray emission angle was near 90º. The flight pass was equal to 176 cm ensuring reliable separation of prompt gamma-rays from neutron and γ-background. The detection of prompt gamma-rays was done using NaI(Tl) scintillator of sizes $\varnothing 2r_d \times h_d \equiv 150 \times 100$ mm surrounded by Geiger-counters. The pulses from NaI(Tl) and Geiger counters were in anti-coincident mode for reducing the influence of cosmic rays background, that enables to make measurements in a wide energy range from 2 MeV and up to ~18 MeV. The full description of experimental set-up and data acquisition system is presented in Ref.[2].

The values of (n,xγ) reaction cross sections $\sigma_\gamma(E_\gamma)$ are unfolded from amplitude spectra by numerical method of approximate solution of ill-posed problem on the compact set [2-4]. Namely, the cross sections $\sigma_\gamma(E_\gamma)$ are obtained from amplitude spectra $S_\gamma(\varepsilon)$ by solving the following integral equation

$$S_\gamma(\varepsilon) = \int_0^{E_{max}} \chi_S(\varepsilon, E_\gamma) \sigma_\gamma(E_\gamma) dE_\gamma , \qquad (1)$$

with $\chi_S(\varepsilon, E_\gamma)$ for the total response function of the spectrometer function with NaI(Tl) response function from Ref.[5], $\varepsilon$ for the γ-ray energy deposited in the detector volume and $E_\gamma$ for the γ-ray energy with maximal value $E_{max}$. This equation is presented then in the matrix form:

$$\vec{S} = X \cdot \vec{\sigma} . \qquad (2)$$

Here, $\vec{S}$, $\vec{\sigma}$ are the vectors with elements $S_i = S_\gamma(\varepsilon_i)$, $\sigma_j = \sigma_\gamma(E_{\gamma,j})$, $X$ is rectangular matrix with elements $X_{i,j} = \chi_S(\varepsilon_i, E_{\gamma,j}) \Delta E_\gamma$ ($i=1,...,m; j=1,...,n$), $m$ is number of bins in the amplitude spectra and $n$ is the number of energy intervals; $\Delta E_\gamma$ ~0.5 MeV is the width of the gamma-ray bins; $E_{\gamma,j}, \varepsilon_{\gamma,i}$ are the values of the corresponding energies in the bins.

The linear algebraic equation (2) was solved by the least squares method to search a minimum of the square of norm of the residual $\vec{r}(\vec{\sigma}) = X \cdot \vec{\sigma} - \vec{S}$,

$$Min\left\{\|\vec{r}(\vec{\sigma})\|^2 = \vec{r}^T \cdot \vec{r} = \left(X \cdot \vec{\sigma} - \vec{S}\right)^T \cdot \left(X \cdot \vec{\sigma} - \vec{S}\right)\right\}, \qquad (3)$$

where symbol "*T*" denotes transpose. For determination $\vec{\sigma}$, the iteration algorithm of the steepest descent method was used in the form

$$\vec{\sigma}_{k+1} = \vec{\sigma}_k - 2\lambda_k \cdot X^T \cdot \vec{r}_k, \quad \vec{r}_k = \vec{r}(\vec{\sigma}_k) = X\vec{\sigma}_k - \vec{S} . \qquad (4)$$

Value of the parameter $\lambda_k$ was found from condition of minimization the square of norm of residual $\vec{r}_{k+1}$,

$$Min\left\{\|\vec{r}_{k+1}\|^2\right\}, \quad \vec{r}_{k+1} = (1 - 2\lambda_k \cdot X \cdot X^T) \cdot \vec{r}_k . \qquad (5)$$

As a zero approximation ($\vec{\sigma}_o$), the results of calculations by Empire code with default input parameters were used. The iteration procedure was stopped at a condition ($\|\vec{\sigma}_{k+1}\|^2 - \|\vec{\sigma}_k\|^2$) $\leq 0.01 \cdot \|\vec{\sigma}_k\|^2$. The uncertainties of unfolded cross sections were evaluated from variation of amplitude spectra in the assumption of Gauss distribution of amplitude spectrum uncertainties due to the large number of independent external factors.

Figure 1 shows the experimental unfolded spectrum and its uncertainties in comparison with the experimental data from EXFOR data base (Refs. [6-10]). Here our results and experimental data from Refs. [6 - 10] are multiplied, like in Ref.[2], by factor $4\pi$, that is, they are considered as the angle integrated gamma-ray spectrum $\sigma_\gamma(E_\gamma) \equiv d\sigma_\gamma(E_\gamma)/dE_\gamma = 4\pi\sigma_\gamma(E_\gamma,\theta_\gamma)$ due to weak dependence of the differential cross sections on gamma-emission angle. Note that we were able to observe γ-transitions for $^{nat}$Sn in the range $E_\gamma \geq 2$ MeV and for $^{nat}$C only from the first excited level with $E_\gamma = 4.43$ MeV due to low intensities of gamma-transitions with other energies.

As one can see from Fig.1, cross section for $^{nat}$Sn smoothly decreases with energy and has nearly constant value in the energy interval 12÷18 MeV, where it was measured for the first time. The value $\sigma_\gamma(E_\gamma = 4.43$ MeV$) = 232 \pm 30$ mb of measured cross section for $^{nat}$C is in good agreement with the results of other authors.

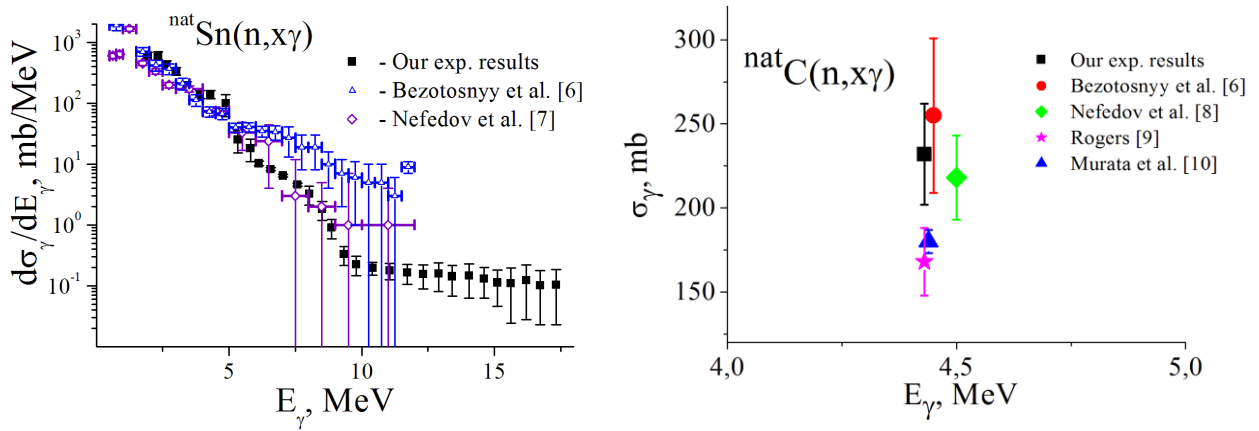

Fig.1. Gamma-ray spectrum from (n,xγ) reactions on $^{nat}$Sn and $^{nat}$C.

**Theoretical calculations and data analysis**

The cross section $\sigma_\gamma(E_\gamma)$ of the target with natural elements are a sum of the cross-sections for each isotope $A_i$, $Z_i$ of the target:

$$\sigma_\gamma(E_\gamma) = \sum_i w_i \cdot \sum_x d\sigma(A_i(n,\{x,\gamma\}))/dE_\gamma, \quad w_i = \alpha_i/100, \quad (6)$$

where $\alpha_i$ is the abundance of the isotope $A_i$, $Z_i$ and $d\sigma(A_i(n,\{x,\gamma\}))/dE_\gamma$ is the differential cross sections of gamma-ray emission from all possible reactions $(n,\{x,\gamma\})$ with any outgoing particle ($x$) and gamma-rays. Calculations were performed using the EMPIRE 3.2 and TALYS 1.6 codes [11, 12] with allowance for outgoing particles and gamma-rays at equilibrium (HF curves in the figures) and from pre-equilibrium states using the default sets of the input parameters, specifically, for EMPIRE code, these parameters are Modified Lorentzian model (MLO1) for the electric dipole photon strength function (PSF) and Enhanced Generalized Super-Fluid Model (EGSM) for nuclear level density (NLD). Calculations were performed with preequilibrium processes (PE) (parameter PCROSS = 1.5) and without PE (PCROSS = 0). For TALYS code, default parameters are Enhanced Generalized Lorentzian (EGLO) for PSF and Gilbert - Cameron approach (GC) for NLD. Global optical potential given by Koning – Delaroche [13] was adopted as default in the calculations within two codes. The following values of abundances were used [14]: $^{116}$Sn (14.3%), $^{117}$Sn (7.61%), $^{118}$Sn (24.03%), $^{119}$Sn (8.58%), $^{120}$Sn (32.85%), $^{122}$Sn (4.72%), $^{124}$Sn (5.94%) for $^{nat}$Sn target, and $^{12}$C (98.93%), $^{13}$C (1.07%) for $^{nat}$C target, where abundance is indicated in brackets.

Figure 2 and Table 1 show comparison of the theoretical calculations with experimental data. One can see that theoretical calculations as a whole are rather good estimates the experimental data. For Sn nuclei, preequilibrium processes are strongly affected emission of gamma-rays of high energies and should be taken into account for gamma-spectrum calculations.

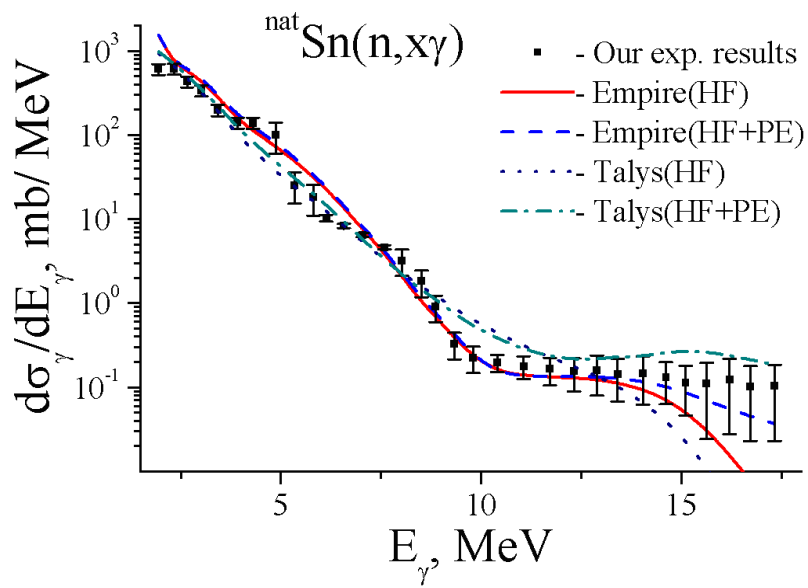

Fig.2. Comparison of experimental data with calculations of gamma-ray spectrum from (n,xγ) reactions on $^{nat}$Sn.

*Table 1.* **Comparison of experimental data with calculations by EMPIRE of inclusive (n,xγ) cross section for $^{nat}$C at $E_\gamma$=4.43MeV**

| Cross-sections | $\sigma_\gamma$, mb |
|---|---|
| Our experimental data | 232±30 |
| Calculations without PE | 204 |
| Calculations with PE | 203 |

Next is considered a sensitivity of the calculated gamma spectra to the PSF and the NLD models; for this aim, we perform the calculations within EMPIRE code with allowance for preequilibrium processes (PCROSS = 1.5). The results are shown in Fig.3 and Table 2. In the calculations of the curves on Fig.3a, the EGSM is taken for nuclear level density and different approaches are used for PSF[14-16]:. Standard Lorentzian model (SLO), MLO1 and MLO4 variants of Modified Lorentzian (MLO) approach, Enhanced Generalized Lorentzian (EGLO) model (according to RIPL-2 ) and Generalized Fermi Liquid (GFL) model. In the results presented in Fig.3b, the model MLO1 is used for PSF and different methods are taken for NLD [14,17]: the Generalized Superfluid model (GSM), Gilbert and Cameron (CG) model, microscopic combinatorial level densities within Hartree-Fock-Bogoliubov method (HFBM) and Modified Generalized Super-Fluid model with Bose attenuated numbers for vibrational enhancement factor (MEGSM).

It can be seen from results in Fig.3, the calculated cross sections are weak dependent on PSF shape except EGLO model for which we obtained much lower values of gamma-spectrum in the 6-15 MeV energy range in comparison with data. The sensitivity of the calculations to the NLD models is much stronger than to the PSF. Considerable disagreement is demonstrated between data and calculated cross sections with HFBM and GC in the energy ranges 5-12 MeV and 3-6 MeV, 10-15 MeV, respectively. Cross section of (n,xγ) reaction on $^{nat}$C for gamma rays of 4.43 MeV has only week dependents on PSF shape.

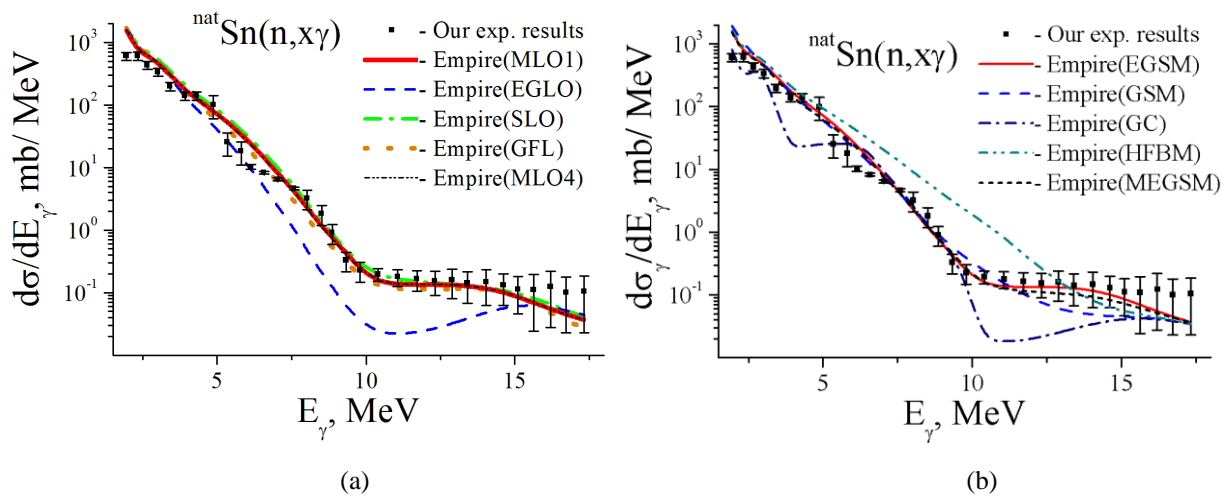

(a)     (b)

Fig.3. Experimental data and corresponding theoretical calculations of (n,xγ) cross sections for $^{nat}$Sn performed using different models for PSF (a) and NLD (b).

*Table2*. **Comparison of experimental data and theoretical calculations of (n,xγ) cross sections for $E_γ$=4.43 MeB $^{nat}$C using different models of PSF and NLD**

| Cross sections | | σ, mb |
|---|---|---|
| Our experimental data | | 232±30 |
| Theoretical calculations | NLD: EGSM<br>PSF: MLO1, EGLO, SLO, GFL, MLO4 | 203 |
| | PSF: MLO1<br>NLD: EGSM, MEGSM | 203 |
| | PSF: MLO1<br>NLD: HFBM, GC | 252 |

## Conclusions

The yield of prompt γ-rays produced by interaction of the 14.1 MeV neutrons with $^{nat}$Sn and $^{nat}$C targets were measured using the time-of-flight technique and cross sections of (n,xγ) reactions were unfolded from amplitude spectra. Determined cross-sections are in a rather good agreement with the data of other experiments [6-10]. For the high energy range (12÷18 MeV), the cross sections of (n,xγ) reactions on $^{nat}$Sn were measured for the first time and corresponding values are approximately equal to 0.10÷0.13 mb/MeV.

Comparisons of the experimental data with the calculations demonstrate high reliability of the calculations with the use of the EMPIRE and TALYS codes [11, 12] with the default sets of the input parameters for estimations of the gamma-ray spectrum in reactions induced by fast neutrons.

This work is supported in part by the IAEA (Vienna) under IAEA Research Contract within CRP No.F41032.